\documentclass[12pt]{article}
\usepackage{amsmath}
\usepackage{amssymb}
\usepackage{subfigure}
\usepackage{psfrag}
\usepackage{epsfig}
\usepackage{graphicx}
\usepackage{tikz}
\usepackage{cite}
\usepackage{xfrac}
\usepackage{hyperref}

\usepackage[numbers,sort&compress]{natbib}

\newcount\xrefpos \xrefpos=0
\newcount\yrefpos \yrefpos=0
\newcount\xput \xput=0
\newcount\yput \yput=0

\def\refpos#1 #2 #3{\global\xrefpos=#1 \global\yrefpos=#2
                         \rlap{$\smash{#3}$}}
\def\put #1 #2 #3{\xput=#1 \yput=#2
                  \advance\xput by -\xrefpos
                  \advance\yput by -\yrefpos
                  \rlap{\kern\the\xput truebp
                        \vbox to 0pt{\vss\hbox{$\displaystyle #3$}
                        \kern\the\yput truebp}}}
\def\beginlabels\refpos#1\endlabels{\hbox{$\refpos#1$}}

\newcommand{\ba}{\begin{eqnarray}}
\newcommand{\ea}{\end{eqnarray}}
\newcommand{\beq}{\begin{equation}}
\newcommand{\eeq}{\end{equation}}

\usepackage{amsfonts,amssymb}

\begin{document}

 \begin{center}
 {\Large \bf Holographic Higgs Phases }

\bigskip
\bigskip
\bigskip
\bigskip

\vspace{3mm}

Moshe Rozali\footnote{email: rozali@phas.ubc.ca}, Darren Smyth\footnote{email: dsmyth@phas.ubc.ca} and Evgeny Sorkin\footnote{email: evgeny@phas.ubc.ca}

\bigskip\medskip
\centerline{\it Department of Physics and Astronomy}
\smallskip\centerline{\it University of British Columbia}
\smallskip\centerline{\it Vancouver, BC V6T 1Z1, Canada}

 \bigskip\bigskip\bigskip

 \end{center}

\abstract{We discuss phases of gauge theories in the holographic context, and formulate a criterion for the existence of a Higgs phase, where the gauge redundancy is ``spontaneously broken'', in purely bulk language. This condition, the existence of a finite tension solitonic string representing a narrow magnetic flux tube, is necessary for a bulk theory to be interpreted as a Higgs phase of a boundary gauge theory. We demonstrate the existence of such solitons in both top-down and bottom-up examples of holographic theories. In particular, we numerically construct new solitonic solutions in AdS black hole background, for various values of the boundary gauge coupling, which are used to demonstrate that the bulk theory models a superconductor, rather than a superfluid. The criterion we find is expected to be useful in finding holographic duals of color superconducting phases of gauge theories at finite density. }

\newpage
\section{Introduction and Conclusions}

Holographic methods are becoming a standard tool in analyzing quantum field theories and revealing physics which would be strongly coupled using other, more conventional descriptions. Most relevant to this note, this basic tool has been utilized in contexts in which we expect ``spontaneous breaking of gauge invariance,'' such as holographic superconductivity (for a review see\cite{Horowitz:2010gk}) or color superconducting phases in QCD (for recent attempts to model such phases see \cite{Chen:2009kx, Basu:2011yg}).

Of course, the expression ``gauge symmetry'' and its breaking is a misnomer, or more precisely relies on specific classical limit for its definition. In a specific weak coupling limit it makes sense to speak of gauge redundancies as approximate global symmetries and use the machinery and language of global symmetry breaking in this context.  However, in an inherently non-perturbative context such as holographic dualities one needs to stick to more precise and gauge-invariant definitions. Such characterization of massive phases of gauge theories was given by 'tHooft \cite{hooft1978phase, hooft1980topological}, and we review this classification in section 2.

This classification of gauge theory phases is gauge invariant and non-perturbative, relying on the response of the gauge theory vacuum to massive external sources. This could be best used in the holographic context whenever we have an idea of the gauge theoretic microscopic definition of the system, and use the holographic context merely to perform calculations in the strongly coupled regime. This situation is demonstrated in section 3, using one particularly simple such ``top-down'' context, namely that of the Coulomb branch of the maximally supersymmetric $SU(N)$ theory in four dimensions. We demonstrate that the phase structure of the theory is manifested in certain geometrical features of the bulk theory which reproduce the expected results.

The purpose of this exercise is to extract a purely bulk criterion for the existence of a Higgs phase interpretation of the theory, which we can then use in situations where the microscopic definition of the bulk theory is less well-understood. Indeed, we see that the expected behavior of the 'tHooft loop operator in the Higgs phase implies the existence of certain type of solitonic strings localized in the IR region of the bulk theory, representing a narrow magnetic flux tube in the boundary theory\footnote{The precise interpretation of this flux tube depends on the microscopic interpretation of the theory, and in particular on the UV region of the geometry.}. In the holographic context, this can be taken as the {\it definition} of such phases, since it implies much of the phenomenology we associate with the Higgs mechanism.

Since our criterion depends only on the bulk geometry, it is ideal in the bottom-up approach to holographic duality, where the microscopic definition of the theory is lacking. In section 4 we demonstrate this criteria in the context of holographic {\it superconductors}, namely holographic theories with the Marolf-Ross prescription \cite{Marolf:2006nd} (see also \cite{Witten:2003ya}) for obtaining boundary dynamical gauge fields (with finite gauge coupling). Such theories, in the broken phase, model genuine superconductors rather than superfluids. We study the bulk and boundary properties of the superconducting vortices, and demonstrate their role in characterizing the phase structure of the holographic theory. 

To this end, we construct new solitonic solutions in $AdS_4$ black hole background (in the probe limit),  for various values of the boundary gauge coupling (the parameter $\alpha$ we introduce in (\ref{coup})), by solving numerically the bulk equation of motion -- a set of coupled non-linear partial differential equation. Section 4 is devoted to setting up the equations and boundary conditions, and describing the properties of the solutions.  Essential to our solutions is the use of dynamical boundary conditions for the gauge fields  (introduced in \cite{Marolf:2006nd}), which are necessary for obtaining finite energy solutions, corresponding to superconducting vortices\footnote{Superfluid vortices were constructed in \cite{Keranen:2009re}. Superconducting vortices, with infinite boundary gauge couplings, were constructed in \cite{Domenech:2010nf}. We compare and contrast our solutions with those solutions below. See also \cite{Montull:2012fy} for a related construction.}. We describe in detail the bulk and boundary properties of our solution, and find a few intriguing  patterns in the dependence of their free energy on temperature and on the boundary gauge coupling.

We are hopeful that the criterion discussed here, and the role it plays in models of holographic superconductivity, will assist in formulating the problem of holographic {\it color} superconductivity, and in constructing holographic models along the lines of \cite{Basu:2011yg}. We hope to return to this problem, one of the original motivations of the present note, in the near future.

\section{Characterization of Gauge Theory Phases}

In \cite{hooft1978phase, hooft1980topological} 'tHooft introduced a classification of phases of gauge theory based on its response to electric and magnetic sources. For the characterization to be a precise definition of the associated phases, we restrict ourselves for now to theories with gauge group $SU(N)/Z_{N}$, such as gauge theories based on unitary groups in which all matter fields are in the adjoint representation. In such theories the centre of the gauge group $Z_{N}$ is a global symmetry which aids in providing order and disorder parameters to characterize the different phases.

The response of the theory to electric sources is measured by the Wilson loop \beq W(C) = Tr(e^{i \int_{C} A})  \eeq where we take the trace in the fundamental representation. The curve C is taken to represent the worldline of two static external sources separated by distance $L$ , and the Wilson line then computes the static potential between these sources.

The response to magnetic sources is similarly represented by a 'tHooft loop $T(C)$, which plays a role of a disorder parameter in the theory. The 'tHooft loop operator is defined in the path integral language as an integral over all gauge field configurations with a prescribed singularity along the curve $C$. The singularity represents the presence of an external magnetic sources. For the curve $C$ which represents the worldline of two well-separated static sources, this operator probes the theory in a way which is similar to the Wilson loop. Indeed, as is well-known, these two observables are exchanged under electric-magnetic duality (see for example \cite{Gomis:2009ir} or section 10 of Witten's lectures in \cite{deligne1999quantum}).

We can then distinguish the different phases\footnote{This is not a complete classification of such phases. For example, there could be critical points and oblique confinement phases distinguished by the behavior of dyonic loop operators. We will not discuss such phases here.} of gauge theories by following asymptotic behavior for large loops $C$:

\begin{itemize}

\item Confinement:   $W(C) \sim e^{-A(C)}$  and $T(C) \sim  e^{-L(C)}$

\item Higgs Phase:  $W(C) \sim e^{-L(C)}$  and $T(C) \sim  e^{-A(C)}$
\end{itemize}
 We denote the area enclosed within the curve $C$  by $A(C)$ and the corresponding behavior of the loop operator is called the {\it area} law. This encodes the linear potential between the corresponding (electric or magnetic) sources. The linear potential has an intuitive picture in terms of the existence of flux tubes connecting the sources (confining strings) which in turn exist because the corresponding (electric or magnetic) flux lines emanating from the sources  form narrow flux tubes and do not spread (the Messiner effect). Similarly, the length of the curve $C$ is denoted by $L(C)$, and the corresponding behavior for the loop operator is called the {\it perimeter} law.  Such behavior encodes the fact that the fields generated by the corresponding source are short ranged (screened) and influence only the close vicinity of the source location. 
 
In the Coulomb phase, or in a conformal field theory, the behavior of both the Wilson and 'tHooft operators is dictated by conformal invariance. For the loops corresponding to static sources separated by distance $L$, we have the behavior  
\beq W(C) \sim T(C) \sim e^{- \frac {a T}{L}} = e^{- T V(L)}  \eeq where $T$ is a large time cutoff, and $a$ is a constant (which can depend on coupling constants of the theory).  Note that while formally this is classified as a perimeter law for both Wilson and 'tHooft loops, the behavior of the static potential $V(L)$  in a massive (screened) phase is different  $V(L) \sim e^{- \frac{L}{L_{0}}}$ where $L_{0}$ is the screening length.

The prescription of calculating the Wilson and 'tHooft loops in AdS/CFT is simple and well-known\footnote{We use in our holographic discussion the BPS loops, the so-called Wilson-Maldacena loop and its magnetic dual \cite{Maldacena:1998im, Rey:1998ik}. The asymptotic behavior for the loops we consider is unaffected by the presence of the scalar fields. For suggestions on calculating the Wilson loop itself see  \cite{Alday:2007he}.}. The expectation value of Wilson and 'tHooft loop respectively, in the fundamental representation and when working in the saddle point approximation, is of the form $e^{-S}$. The action $S$ is the minimal action of the worldsheet of fundamental string or D-strings respectively, in a configuration which end on the prescribed curve $C$ on the boundary. As quantum operators the Wilson and 'tHooft loops obey an interesting algebra which constrains the possible phases of gauge theory, which was discussed in the context of AdS/CFT by Witten (section 5 of  \cite{Witten:1998wy}). 

The qualitative behavior of the Wilson loops in confining theories is also well-known. The electric flux tube connecting external sources is mapped into a string worldsheet dipping into the bulk. In the confining phase the electric flux lines are confined to narrow flux tubes. The dual statement is that the string worldsheet localizes in the bulk radial direction, oftentimes for clear geometrical reasons (e.g. the ``end'' of the IR geometry in some sense). In the next two sections we provide an analogous statement for magnetic flux tubes in holographic theories in the Higgs phase\footnote{For previous discussion of these flux tubes, see  \cite{HoyosBadajoz:2008fw}.}.

\section{Top Down Model}

Consider  $k$ flat probe D3 branes in $AdS_{5}\times S^{5}$ located at $r=v$ (in Poincare coordinates), and smeared over the sphere $S^{5}$. Here $v$ is proportional to the VEV of the adjoint Higgs field giving rise to the Higgs mechanism in the $\mathcal{N}=4$ SYM theory. This corresponds to the pattern of symmetry breaking $SU(N)\rightarrow SU(N-k) \times SU(k)$, in the large $N$ limit, while $k$ is kept finite\footnote{This is conventionally called to Coulomb phase, and indeed the leading order interaction between electric sources will be Coulomb-like. Nevertheless we'll use the term Higgs or broken phase.}. 
In this example we have the power of large $N$ as an organizing principle, and we'll see that it aids us in separating the effects of symmetry breaking on the electric and magnetic loop operators.  

\subsubsection*{Electric Flux Lines}

We are mainly interested in magnetic flux tubes, but we start with a brief discussion of the Wilson loop. In the broken phase, with the above breaking pattern, the static potential between electric sources in the fundamental representation is schematically of the form \beq \label{electric} V(L) = \frac{a}{L} + b \, \frac{k}{N} \,\frac{e^{- c v L}}{L} \nonumber \eeq where $a,b,c$ are constants. The leading order potential is still Coloumb-like, but since $k$ gauge bosons are now massive we have a $\frac{1}{N}$ correction involving exchange of those massive gauge bosons. This can be seen, for example, if we repeat the calculations of \cite{Erickson:1999qv, Erickson:2000af} in the broken phase. 

In the bulk this modification can be explained simply, as follows. The Wilson loop calculation corresponds, in the saddle point approximation, to finding the area of a fundamental string worldsheet whose boundary ends on the prescribed curve $C$. The leading order term  in the $\frac{1}{N}$ expansion contributing to (\ref{electric}) corresponds to the calculation in pure AdS\cite{Maldacena:1998im, Rey:1998ik}. The form of the leading $\frac{1}{N}$ correction in (\ref{electric}) suggests a modification of the action of the same saddle point.

\begin{figure}[h!] 
  \centering
    \includegraphics{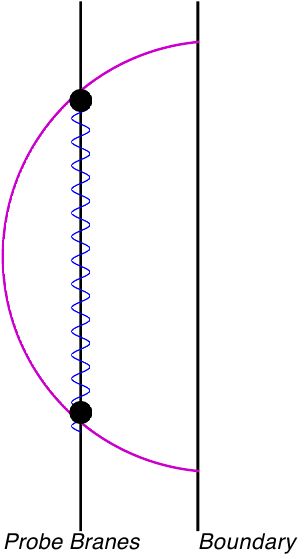}
    \caption[]{Wilson loop stretched between widely separated sources on the boundary. The leading order correction in the $\frac{1}{N}$ expansion comes from exchange of massive scalar representing the radial fluctuations of the probe branes.}
\label{electricW}
\end{figure}

The required modification arises when considering the worldvolume theory on the probe D3 branes. Consider the worldsheet of the fundamental string for well separated electric sources, in the broken phase, a situation which is depicted in figure \ref{electricW}. The leading order contribution for the Wilson line corresponds to the worldsheet area, and  $\frac{1}{N}$  corrections come from interactions between the part of the worldvolume intersecting the probe branes (two lines on the probe branes,  represented by two points in figure \ref{electricW}) . Since the radial fluctuations of the probe branes are massive -- those correspond to the longitudinal modes of the W-bosons -- it is easy to see that exchange of the massive scalar fields corresponding to these brane fluctuations reproduce the form of the leading $\frac{1}{N}$ correction in figure \ref{electricW}.

\subsubsection*{Magnetic Flux Lines}

In contrast to the calculation of the Wilson loop outlined above, the `tHooft loop  expectation value changes character from perimeter to area law, already in the leading order in the $\frac{1}{N}$ expansion. This corresponds to the existence of a new type of saddle point, rather than a modification of the action of the existing worldsheet. 

The new saddle point is similar to that of the Wilson loop in {\it confining} theories. Indeed, in such case the geometry of the bulk provides an IR cutoff, such as a soft or hard wall, or cap to the geometry. The area law is realized geometrically as the Wilson line for widely separated electric sources receives contributions predominantly from the vicinity of the IR geometry. This is the holographic dual to the statement that the flux lines connecting two electric sources do not spread out in the confining vacuum.

\begin{figure}[t!] 
  \centering
    \includegraphics{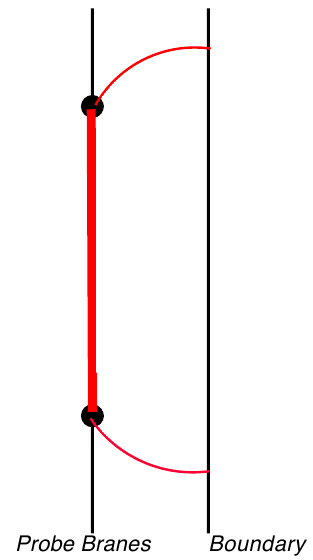}
    \caption[]{World volume of D1 brane stretched between widely separated magnetic sources on the boundary. The area law for the 'tHooft loop results from the existence of a string-like object localized in the radial direction. In this model such object can be represented as soliton on the worldvolume of the probe branes, drawn in a thick red line along the worldvolume of the probe brane.}
\label{magneticT}
\end{figure}

In our case the magnetic dual to that statement cannot be explained in terms of the bulk geometry alone. Indeed, the area law for the D1 brane has to arise from differences between the worldvolume theory of such brane and that of a fundamental string (for example the different dilaton coupling \cite{HoyosBadajoz:2008fw}). In our simple model this is easy to identify: in the presence of the probe branes the worldvolume of the D1 branes can take a ÒdetourÓ through the probe D-branes which, for widely separated magnetic sources, will minimize the action. This is due to the fact that on the worldvolume of the probe branes, the D1 brane can be transformed into a solitonic string of finite tension. Therefore asymptotically in such separation, the minimum action configuration would be the one depicted in figure \ref{magneticT} in which the D-string worldvolume stretches mostly along the worldvolume of the probe branes. Note that this is qualitatively similar to the Wilson line in a confining theory, in that the radial location of the loop is stabilized at some fixed radial location for widely separated sources.

The interpretation of the flux lines in this simple example depends in various ways on understanding the full gauge-gravity duality. In particular, we have used large $N$ scaling to distinguish electric from magnetic flux tubes, and correspondingly confinement from the Higgs mechanism. Furthermore, the microscopic interpretation of the theory helped identify the type of charges available in the gauge theory, and which can be connected by those flux tubes. 
Nevertheless, we have identified a necessary condition for the existence of Higgs phase interpretation of the theory: the bulk spacetime should support a finite tension solitonic object which is approximately localized in the radial direction. The existence of this object, dual to a narrow magnetic flux tube, is necessary for the `tHooft loop of the boundary theory to obey an area law. In the next section we demonstrate the existence of such solutions in a simple bottom-up model of holographic superconductivity.

\section{Application to Holographic Superconductivity}
In this section we discuss a specific 2+1 dimensional bottom-up model of holographic superconductivity \cite{Hartnoll:2008vx}. As argued above, an area law for the 'tHooft loop is guaranteed by a finite energy vortex solution of the bulk fields localized in the radial direction, representing magnetic flux tube in the boundary theory.  When we do not have a microscopic definition of the theory, we take the existence of  such soliton as the {\it definition} of the Higgs phase in the bottom-up holographic context. We demonstrate below the existence of such finite energy solitons in the present context.

Crucial to the analysis is the prescription given in \cite{Marolf:2006nd} (see also \cite{Witten:2003ya}) for obtaining dynamical gauge fields in the boundary theory, by requiring the bulk gauge fields to obey a specific type of  boundary conditions in the UV, which we will refer to as ``dynamical`` boundary conditions. We show that with these boundary conditions the required vortex solutions exist, and furthermore have finite energy per unit length. This indicates that the model, in the broken phase, describes a genuine superconductor. 

We then discuss the bulk and boundary properties of the solutions, including the dependence of their tension on the temperature and the boundary gauge coupling.  Finally, the Higgs phase is characterized by electric screening, which we demonstrate by examining the two point function of the boundary gauge field. 

\subsubsection*{The Model}
We work in the context of the bottom-up model of  \cite{Hartnoll:2008vx}. The action is:
\begin{align}
\label{Saction}
& S  =\frac{1}{2 \kappa} \int d^4 x \sqrt{-g} \left [ R-\frac{1}{4} F^{2}_{\mu \nu}-|(\partial_{\mu}-i q A_{\mu}) \psi |^2 -V(\psi,\psi^* ) \right ] \nonumber \\
& V(\psi,\psi^*) =\frac{6}{L^2} + m^2 \psi \psi^*
\end{align}
 where $ m^2 < 0$, and $q$ is the charge of the scalar field. We work in the probe limit   \cite{Hartnoll:2008vx}, defined as:
 \beq
 q\propto \epsilon^{-1}~~~ ~~A_{\mu},\phi\propto \epsilon~~~~~\epsilon \rightarrow 0
 \eeq 
 In this limit the matter energy momentum tensor scales as $\epsilon^{2}$ and drops out of the Einstein equation, and the metric is unaffected by the matter fields, while the Maxwell and scalar equations remain unchanged. The background solution features an AdS Schwarzschild black hole geometry which, for certain values of the thermodynamic variables (the chemical potential $\mu$, or equivalently the temperature $T$) develops a profile for the scalar condensate and the temporal component of the gauge field. This signals the onset of symmetry breaking below the critical temperature.

 We choose to work in cylindrical coordinates $(t,\rho,\theta,w)$ with the conformal boundary located at $w=0$. Our background metric is then:
 \begin{align}
ds^2= \frac{L^2}{w^2} (-f(w)dt^2 + f(w)^{-1} dw^2 +d \rho^2 + \rho^2 d \theta^2)
\end{align}
 In addition we make use of the scaling symmetries of our action to scale the horizon to $w_{+}=1$ and we take the AdS radius of curvature to be  $L=1$. This fixes the metric function to be $f(w)={1}/{w^2}-w$. By dimensional analysis we expect all physical quantities to be proportional to the ratio of ${T}/{\mu}$. In what follows we fix $\mu=1$ and examine the behavior of the various quantities as a function of $T$.
  
\subsubsection*{Ansatz and boundary conditions} \label{bound_cond}
We now discuss matter excitations to the homogeneous background. Guided by the known vortex solutions of the Abelian Higgs model in flat spacetime (reviewed in appendix A), we propose the following ansatz for the solutions we seek:
\begin{align}
A_\mu&\rightarrow (A_0(w,\rho),0,A_{\theta}(w,\rho),0) \nonumber \\ \psi &\rightarrow \psi(w,\rho) \exp(i s \theta)
\end{align}
where  $s$ is the topological number associated with the vortex solution\footnote{Of course, this number is not conserved in the full geometry, and indeed as we will see it ``unwinds`` as function of the bulk radial coordinate $w$.}. 

The equations of motion consist of the two Maxwell and one scalar equation:
\begin{align}\label{PDEs}
& R^2 \left(\frac{q s}{w^2 f}-\frac{q^2 \text{$A_\theta $}}{w^2 f}\right) + \left(\frac{f'}{f}+\frac{2}{w}\right) {\partial_w A_\theta}-\frac{ \partial_\rho A_\theta }{w^2 \rho f}+\frac{\partial^2_\rho A_\theta}{w^2 f}+\partial^2_wA_\theta =0 \nonumber \\
& -\frac{q^2 A_0 R^2}{w^2 f}+\frac{\partial_\rho A_0}{w^2 \rho  f}+\frac{\partial^2_\rho A_0}{w^2 f}+\partial^2_w A_0 =0 \nonumber \\
& R \left(\frac{q^2 \text{$A_0$}^2}{w^4 f^2}-\frac{(s-q \text{$A_\theta $})^2}{w^2 \rho ^2 f}+\frac{f'}{wf}-\frac{m^2}{w^4 f}\right)+\left(\frac{f'}{f}+\frac{2}{w}\right) \partial_w R +\frac{\partial_\rho R }{w^2 \rho f}+\frac{\partial^2_\rho R}{w^2 f}+\partial^2_w R=0
\end{align}
where we rescaled the scalar field as $\psi(w,\rho) \rightarrow w R(w,\rho)$, for reasons of numerical stability. It can be seen that in the probe limit described above there is a scaling symmetry of the equations (\ref{PDEs}), implying that if a solution is found for a given value of $q$ it is known for all $q$ via an appropriate rescaling of the fields. This property is convenient for numerical purposes as it allows us to choose a scale for the matter fields which is numerically tractable.

We wish to solve our system of PDEs on the domain defined by $w_0\leq w\leq w_{+}$ and $0\leq \rho \leq \infty$, where $w_{0}$ is a UV cutoff. For the problem to be well posed we must choose self consistent boundary conditions which are also compatible with the bulk equations of motion. We choose the following boundary conditions on the four different segments of the boundary:

\begin{itemize}
  \item $ \rho \rightarrow \infty $: In flat space it is known that the vortex fields decay exponentially towards asymptotic values for the gauge and scalar fields as $\rho$ goes to infinity. Anticipating similar behavior, in our numerical implementation we impose a Neumann boundary conditions at some finite and large value, $\rho_{{cut}}$, since in that region the solution should tend to the homogeneous ground state\footnote{$\rho_{cut}$ is chosen such that our solutions vary by less than $0.01\%$ if it is increased.}.
  
\item $ \rho \rightarrow 0 $: To determine the boundary conditions at the vortex core we require that all components of the bulk magnetic field be finite. The radial and transverse components of the magnetic field are given by $B_w= {\partial_{\rho} A_\theta}/{\rho}$, and $B_{\rho}={\partial_{w} A_\theta}/{\rho}$, respectively. Finiteness of the radial component implies that $\partial_{\rho}A_\theta \rightarrow 0 $ as $\rho \rightarrow 0$. Regularity of the transverse component then restricts the $A_{\theta}$ component to obey (in this limit) $\partial_w A_{\theta} \rightarrow 0$. Therefore we conclude that $A_{\theta}(w,\rho=0)$ must be a constant. If  we were to impose a Dirichlet conditions at the conformal boundary this would fix this constant to be zero. In our case we have a residual gauge freedom\footnote{The dynamical boundary conditions at the conformal boundary allow for gauge transformations whose parameter is independent of the radial coordinate $w$.}, consistent with the boundary conditions, which we use  to set $A_{\theta}(w,\rho=0)= 0$ at the core of the soliton. 
\item $ w \rightarrow w_0$ : On the conformal boundary we impose Dirichlet conditions on the scalar and $A_0$ fields --- the scalar field must be normalizable and $A_0$ must asymptote to the chemical potential $\mu$. Crucially, on the $A_\theta$ field we impose the following boundary condition
\beq  \frac {\partial A_{\theta}}{\partial w} 
 = \alpha  \, \rho \frac {\partial}{\partial \rho} (\frac{1}{\rho} \frac {\partial A_{\theta}}{\partial \rho} )\label{coup}\eeq
 at the conformal boundary. This corresponds to having a theory in which the boundary value of the bulk gauge field corresponds to a gauge field \cite{Marolf:2006nd} in the boundary theory\footnote{We choose to make dynamical only the component $A_{\theta}$ of the gauge field, for the sake of simplicity, we do not expect the features of the solution to change much if $A_{0}$ is made dynamical as well, since it already has nearly vanishing radial derivative near the conformal boundary, in all the solutions we are interested in.}. The parameter $\alpha$ determines the gauge coupling $e^{2}$ of the boundary gauge field, $e^{2} = g_{bulk}^{2}/ \alpha$. Indeed, having a consistent variational principle requires the addition of the boundary action to \ref{Saction} :\beq S_{bdy}= \frac{1}{e^{2}}\int d^{3}x \, \sqrt{-h} \, F^{2} \eeq where the integration is over the boundary whose induced metric is denoted by h. F is the field strength for the boundary gauge field. We refer to these boundary conditions as the ``dynamical'' boundary conditions in what follows.

\item $ w \rightarrow w_{+} $: Regularity conditions at the horizon are necessary since the equations degenerate there. Choosing the solutions which are regular at the horizon means that the coefficients of the divergent terms in a power series expansion of the equations near the horizon have to vanish. This prescription yields the following constraints in our case:
\begin{align}\label{hor_con}
&A_{0}=0 \nonumber \\
& R \left(-\frac{q^2 \text{$A_\theta $}^2}{\rho ^2}+\frac{2 q s \text{$A_\theta $}}{\rho ^2}-m^2-\frac{s^2}{\rho^2}-3\right)-3 \partial_w R +\frac{\partial_\rho R}{\rho }+\partial^2_\rho R =0  \\
& R^2 q\,\left(s-q \text{$A_\theta $}\right)-3 \partial_w R -\frac{\partial_\rho A_\theta}{\rho }+\partial^2_\rho A_\theta=0 \nonumber
\end{align}
\end{itemize}

We numerically solve the equations with these boundary conditions using successive overrelaxation (SOR) in the domain $w_0\leq w \leq w_+$ and $0 \leq \rho \leq \rho_{cut}$, where the truncation radius $\rho_{cut}$ is large but finite.    In this approach the equations are discretized on a lattice covering the domain of integration. We use a second order finite differencing approximation in which the derivatives are replaced with their finite differencing counterparts.  An initial guess for the value of the scalar and gauge fields is then assigned to each grid point. Dirichlet boundary conditions are implemented by insisting that the initial values assigned to the fields at the boundary grid points are maintained throughout the relaxation procedure, whereas Neumann or Robin boundary conditions must be imposed after each iteration. This is done by using the discrete form of the derivative operators to update the boundary grid points based on the values calculated for the interior points. The SOR algorithm then provides an iterative method of finding numerical solutions to this finite difference system to within a prescribed tolerance. Once the solutions are available other quantities of interest such as the energy density are calculated via insertion of these solutions into the suitably discretized action. Further details of our implementation are found in Appendix B.

\subsubsection*{Free energy}
Before presenting the numerical solution and discussing itÕs properties, we explain the reason we expect the energy (per unit length) to be finite in our case. The discussion parallels that of \cite{Keranen:2009re}.

The Lagrangian density of the bulk fields is:
\begin{eqnarray}\label{actions}
L = & \frac{q^2 \rho  \text{$A_0$}^2 R^2}{2 f} - R^2 \left(\frac{(s-q \text{$A_\theta $})^2}{2 \rho } +\frac{1}{2} w^2 \rho f+\frac{m^2 \rho }{2}\right)-\frac{1}{2} w^4 \rho  f (\partial_w R )^2-w^3 \rho  f R \partial_w R -\frac{1}{2} w^2 \rho (\partial_\rho R )^2  \nonumber\\  &+\frac{\rho  (\partial_\rho A_0)^2}{2 w^2 f}+\frac{1}{2} \rho  (\partial_w A_0)^2-\frac{w^2 f (\partial_w A_\theta)^2}{2 \rho}-\frac{(\partial_\rho A_\theta )^2}{2 \rho }
\end{eqnarray}

	Since the resulting action diverges near the boundary, we regularize it by subtracting the action of the translationally invariant hairy black hole solutions from the on-shell vortex action. Such subtraction automatically removes the divergences which occur due to integration in the $w$ direction. Therefore divergences, if they exist, can occur only as a result of $\rho$ integration. In the region of large $\rho$ the scalar and $A_0$ fields asymptote to their values  in the translationally invariant ground state. Therefore the only terms in the (asymptotic) Lagrangian density to survive the subtraction procedure are:
\begin{align}\label{asympterm}
&-2 \pi \int dt \int_{0}^{1} dw \int_{0}^{\rho_{cut}} d \rho \left[\frac{R^2 (s-q \text{$A_\theta $})^2}{2 \rho } + \frac{w^2 f (\partial_w A_\theta)^2}{2 \rho } \right]
\end{align}
where all fields are understood to be functions of $w$ only. Here we have introduced the cutoff $\rho_{cut}$ in order to regulate potential divergences in the $\rho$ integration. 

If we now use the $A_\theta$ equation of motion to make the substitution:
\begin{align}
-\frac{1}{2} R^2 (s-q \text{$A_\theta $})=\frac{w \left(w f'+2 f\right) \partial_w A_\theta}{2 q}+\frac{w^2 f\partial^2_w A_\theta}{2 q}
\end{align}
and integrate by parts, using the fact that $f$ vanishes on the horizon, we obtain the logarithmically divergent term:
\begin{align}\label{finite}
 & \pi \log{\left(\frac{\rho_{cut}}{\gamma}\right)} \int dt \left(w^2 f \text{$A_\theta $} \partial_w A_\theta \mid_{w=0} + \left(\frac{s}{q}\right) \int_{0}^{1} dw \partial_w (w^2 f \partial_{w} A_\theta) \right) \nonumber \\
 & = \pi \log{\left(\frac{\rho_{cut}}{\gamma}\right)} \int dt \left(w^2 f \text{$A_\theta $} \partial_w A_\theta -\frac{s}{q} w^2 f \partial_{w} A_\theta \right) \mid_{w=0}
\end{align}
In integrating by parts we have introduced the length scale $\gamma$ which is a measure of the size of the vortex core. 

This reasoning led the authors of \cite{Keranen:2009re} to conclude that their vortex solution is logarithmically divergent, as expected from vortices in a superfluid. We see that if we instead consider dynamical boundary conditions for the $A_\theta$ field,  then $\partial_{w} A_{\theta}|_{w=0}=0$ outside the core of the soliton. Then, provided an appropriate vortex solution exists, the coefficient of the logarithmic divergence will vanish. We see below that indeed such vortex solutions (whose profile significantly differs from the superfluid vortices found in \cite{Keranen:2009re}) do exist and we calculate their finite energy (per unit length). This demonstrates that our model describes a genuine superconductor\footnote{This was shown for $\alpha=0$ in \cite{Domenech:2010nf}.}.

\begin{figure}[t!] 
  \centering
    \includegraphics[scale=0.5]{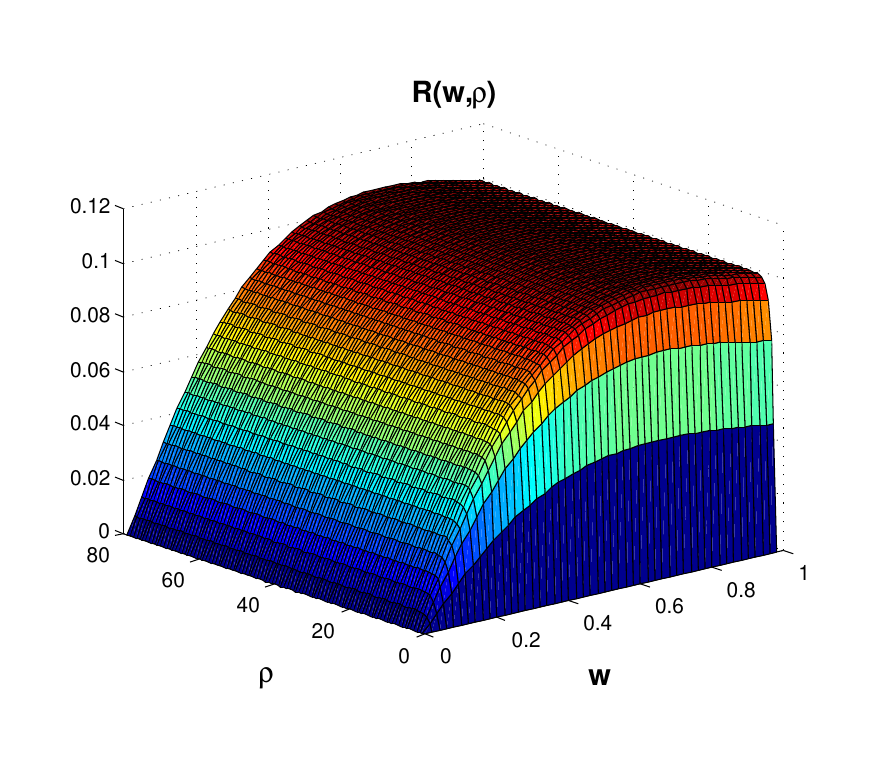}
   \includegraphics[scale=0.5]{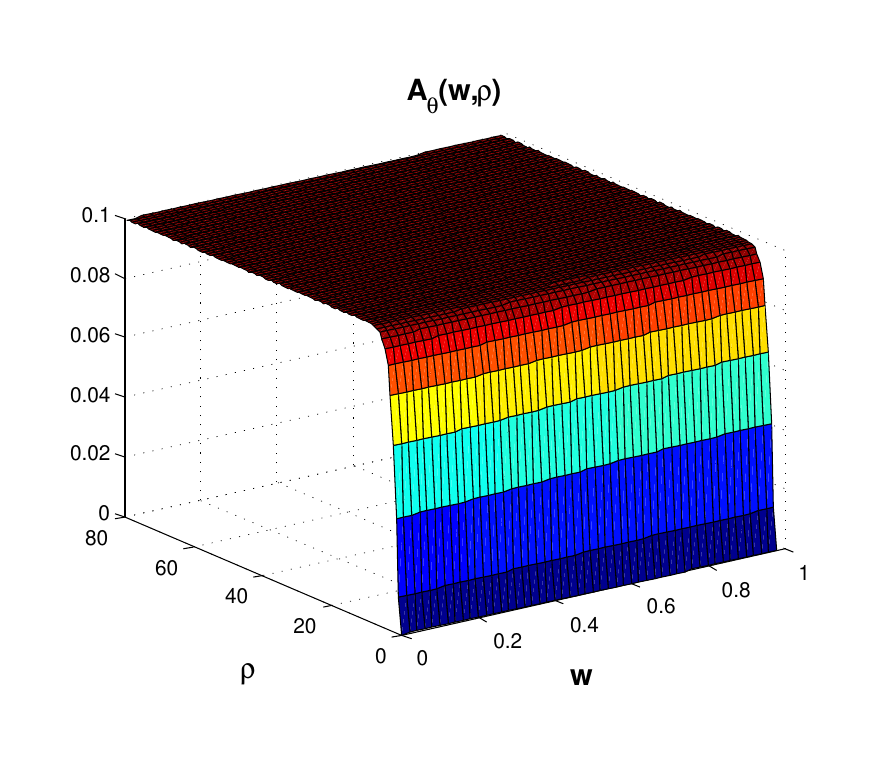}
    \includegraphics[scale=0.5]{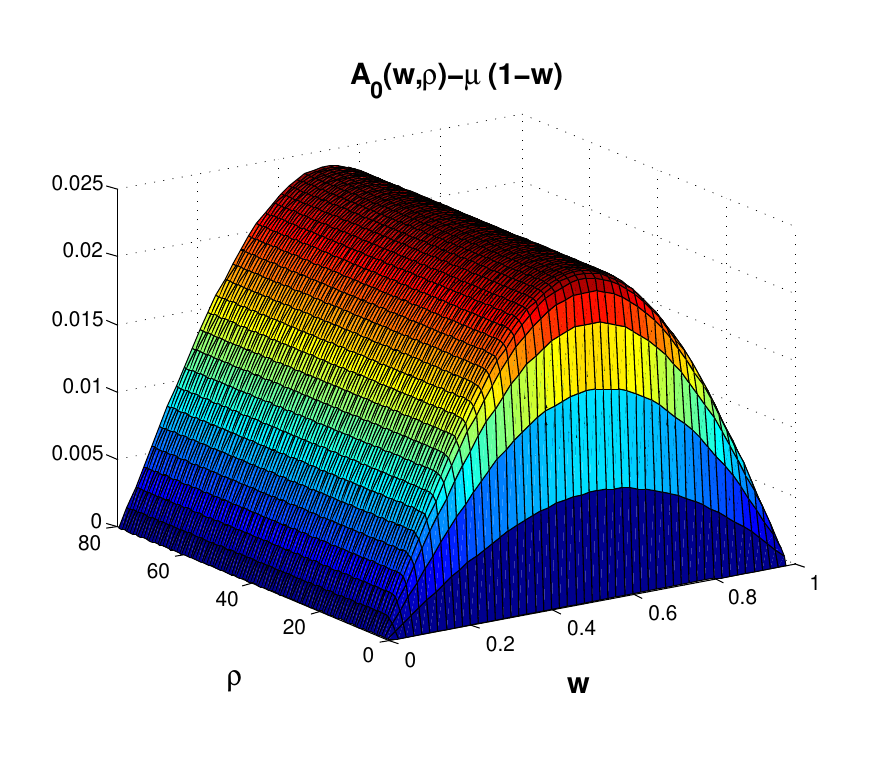}
    \caption{The matter field profiles at a temperature of $\simeq 0.89 T_c$  In order to aid in visualization the background translationally invariant solution has been subtracted from the $A_0$ gauge field. Note the asymptotic approach of the scalar and $A_0$ fields to their translationally invariant profiles and the fact that $A_\theta$ field is independent of the radial coordinate $w$, and asymptotes to $\frac{s}{q}$ as $\rho \rightarrow \infty$.}
\label{fields}
\end{figure}

\subsubsection*{The Solutions - Bulk Properties}

We are now ready to present our numerical solutions for the bulk fields and discuss their properties for different values of the parameter $\alpha$. We leave discussion of our numerical solution to appendix B.

The system has a critical temperature $T_{c}$, below which it is in the condensed phase (i.e. the scalar field develops a normalizable background). Below that critical temperature vortex solutions start appearing, in figure \ref{Bulk_energy}  we show the profile of the fields for a typical vortex solution for $\alpha=0$. The form of the solutions may be understood as follows:  far from the vortex core the fields tend to their homogeneous profiles and the PDEs reduce to ODEs. When solving these ODEs numerically one finds that the solution for the $A_\theta$ field is a constant, given by $\frac{s}{q}$. Together with our previous discussion of the $\rho \rightarrow 0$ boundary conditions, this means that $A_\theta$ asymptotes to a constant, independent of the radial coordinate, both as $\rho \rightarrow 0$ and as $\rho \rightarrow \infty$. Since we are also demanding vanishing radial derivative at the conformal boundary, a reasonable guess is that the global $A_\theta$ solution depends only on $\rho$, i.e. $A_\theta (w, \rho) \rightarrow A_\theta (\rho)$. This is indeed what we find numerically. As seen in (\ref{finite}) above, the asymptotic form of $A_{\theta}$ is directly responsible for the finiteness of the vortex energy.

\begin{figure}[t!] 
  \centering
    \includegraphics[scale=0.5]{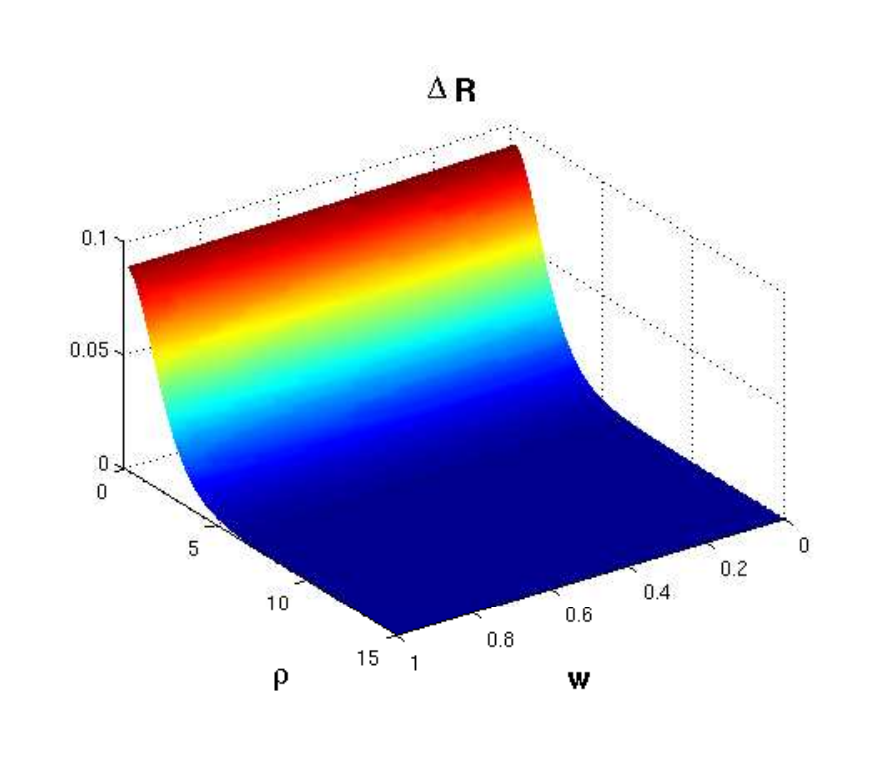}
   \includegraphics[scale=0.5]{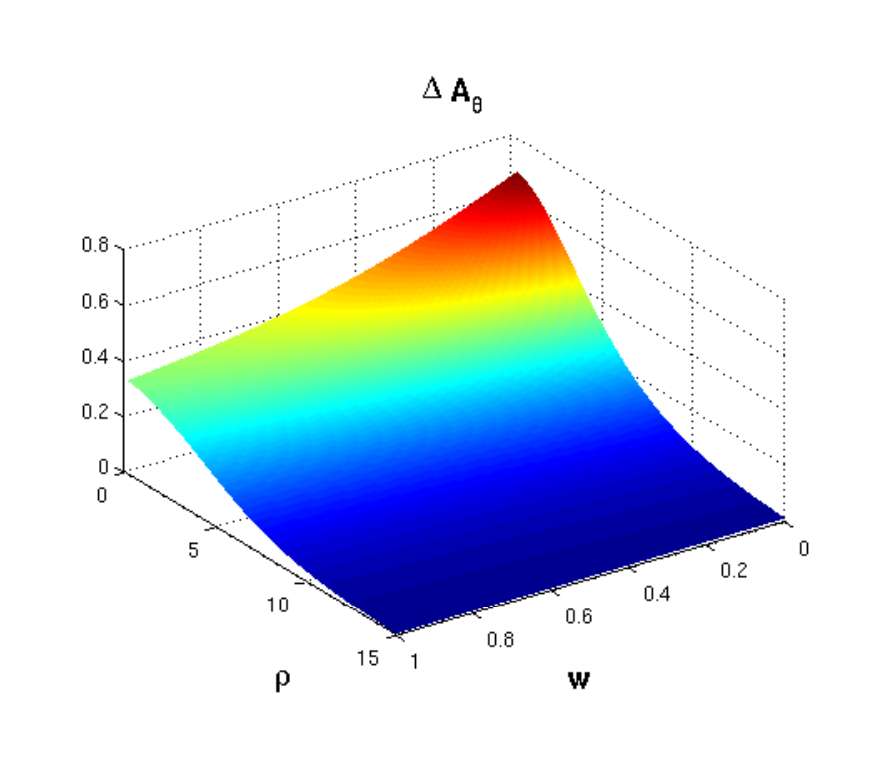}
    \includegraphics[scale=0.5]{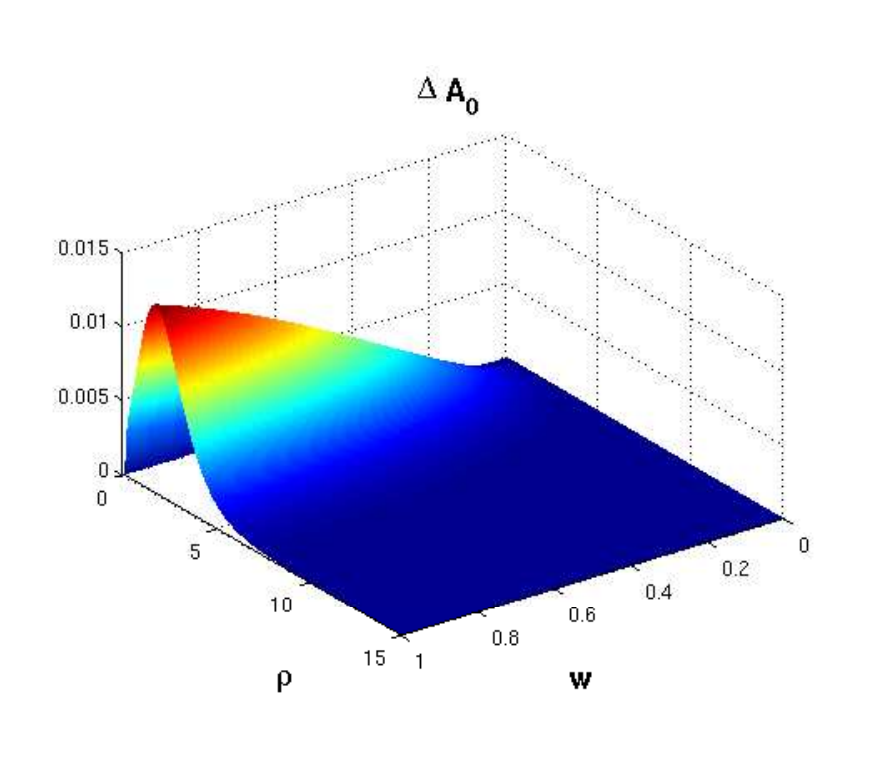}
    \caption{The additional contribution to the bulk fields resulting from the addition of a boundary action with $\alpha=3$. The plots are normalized with repect to the $\alpha=0$ profiles. It can be seen that, as expected, the greatest variation is seen in the $A_{\theta}$ field.}
\label{fields_{diff}}
\end{figure}

\begin{figure}[t!]
	\centering
   \includegraphics[scale=0.5]{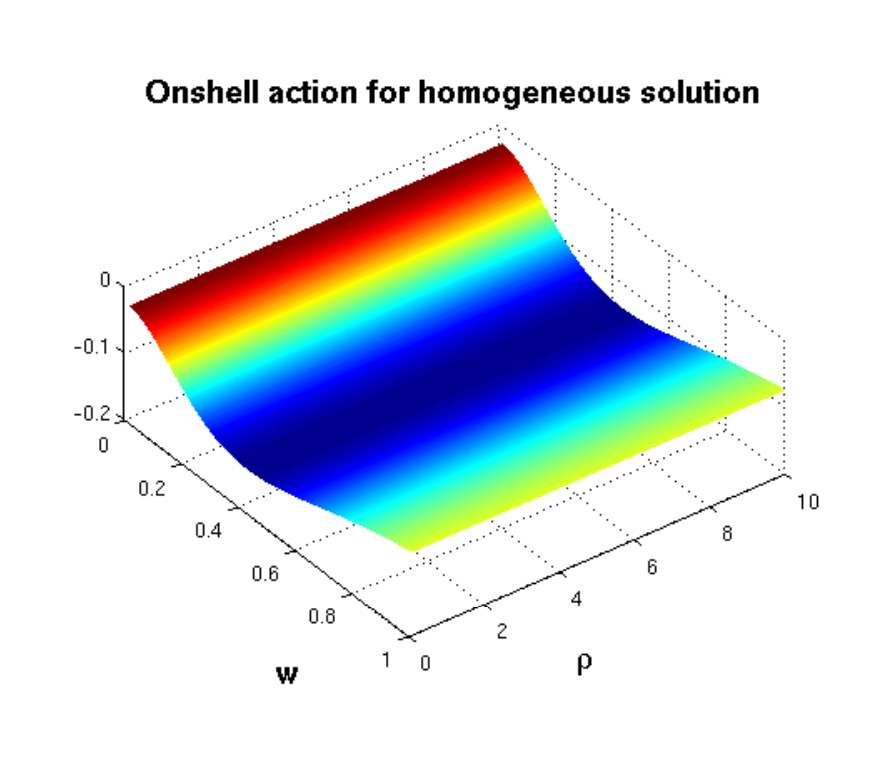}
    \includegraphics[scale=0.5]{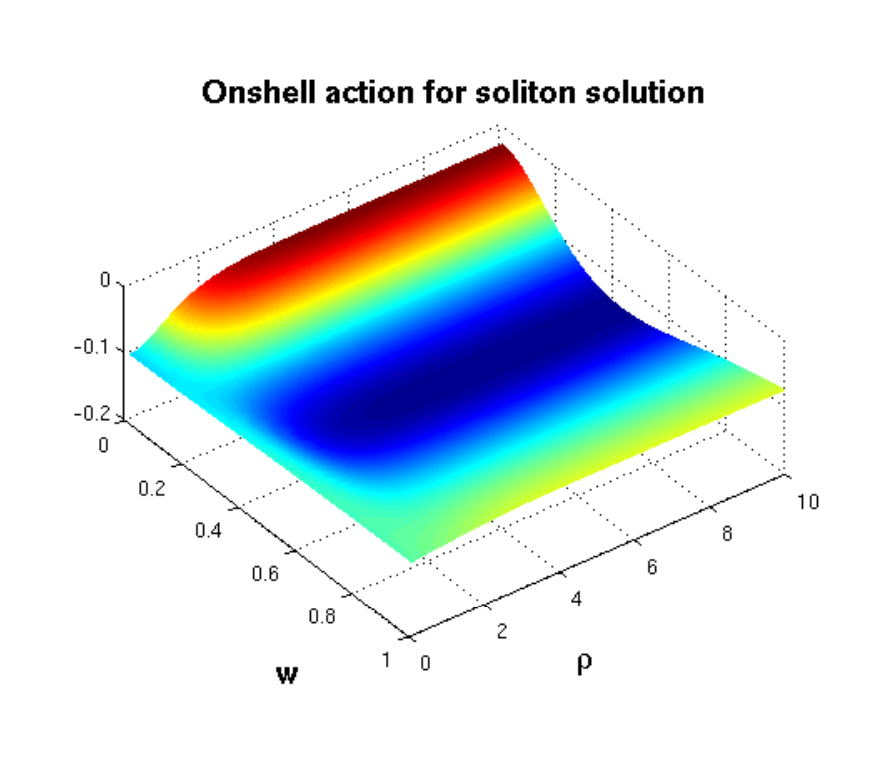}
    \includegraphics[scale=0.5]{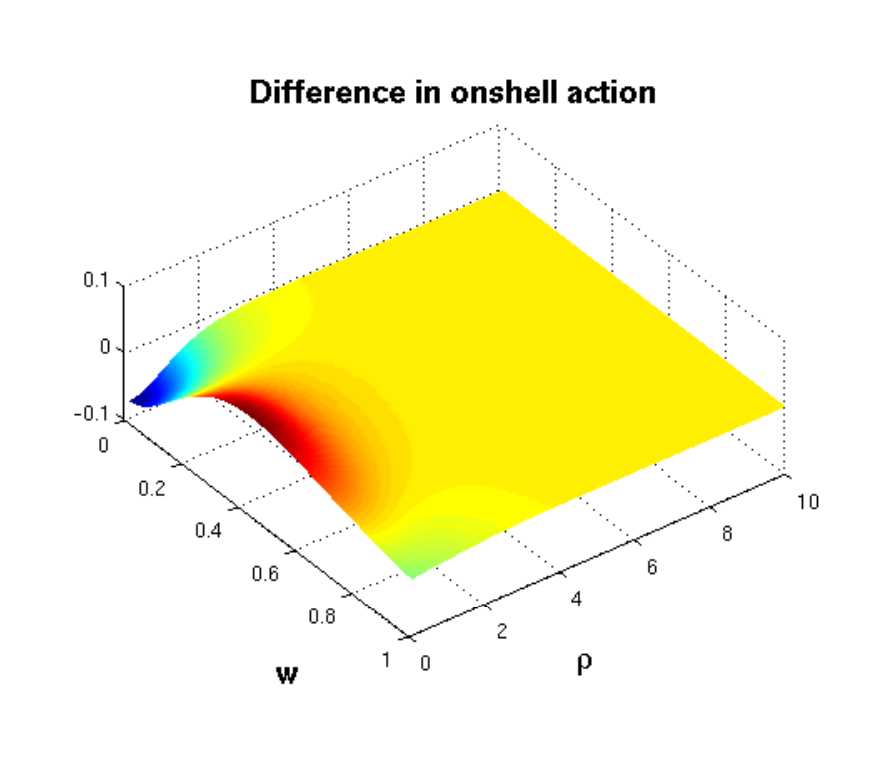}
   \caption{Free energy density profiles for the translationally invariant background and vortex solution, and their difference for $\alpha=0$ and a temperature of $\simeq 0.91 T_c$. Note for ease of visualization we have included the $w$ factors coming from the measure.}
\label{Bulk_energy}
 \end{figure}
 
We have also obtained the solution with $\alpha \neq 0$, in other words with dynamical boundary gauge fields. In figure \ref{fields_{diff}} we demonstrate the effect of the boundary action by displaying the differences in bulk fields (relative to the $\alpha=0$ case) for the specific case of $\alpha=3$. It can be seen that the profile of the fields is no longer homogeneous in the $w$ direction near the core of the vortex. The greatest inhomogeneity is seen in the $A_{\theta}$ and $A_{0}$ fields while the changes in the $R$ field, while substantial in magnitude, are largely homogeneous in $w$.

\begin{figure}[t!]
	\centering
   \includegraphics[scale=0.5]{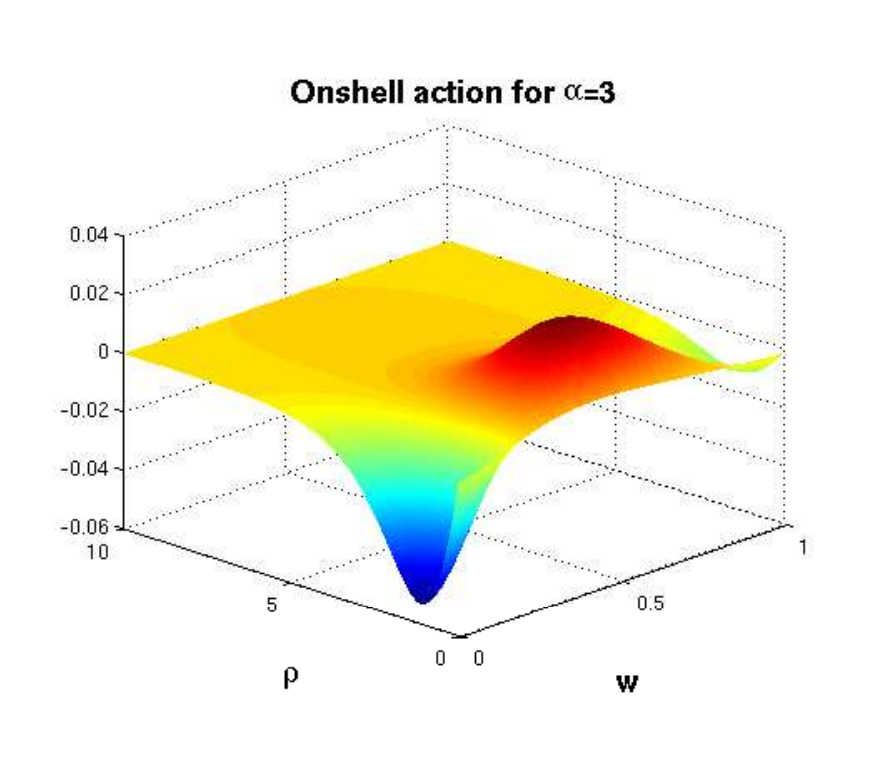}
   \includegraphics[scale=0.5]{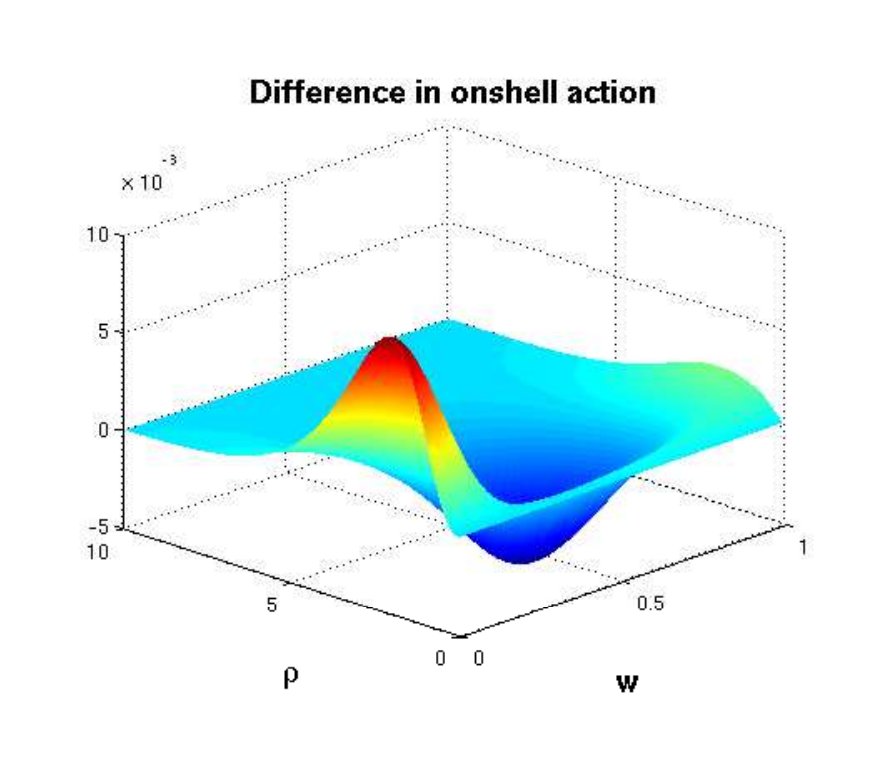}
 \caption{Bulk free energy for the vortex solution with $\alpha=0$, and the difference between this and the $\alpha=3$ solution at a temperature of $\simeq 0.91 T_c$. We note the increased energy density near the conformal boundary relative to the $\alpha=0$ case.}
\end{figure}

Once we obtain the numerical solutions for the matter fields, their on-shell action can be evaluated.  In figure 5 we illustrate the profile of the free energy density of both the translationally invariant and vortex solutions for $\alpha=0$, and their difference. We note that the bulk free energy density of the vortex solution, in the vicinity of the core of the soliton, dips below that of the homogeneous ground state near the conformal boundary. Nevertheless, as we will see below the boundary free energy density of the vortex  (relative to the background) is everywhere positive.

We next turn to solutions with $\alpha \neq 0$. In figure 6 we plot the profile of the bulk energy density (with the homogeneous background subtracted) for the $\alpha=3$ solution, and the difference between that solution and the $\alpha=0$ solution. We see that the change in the free energy density can be significant and is heavily localized near the conformal boundary and the core of the vortex. We also note that, as expected, increasing $\alpha$ has the effect of shifting more of the contribution of the action to the vicinity of the conformal boundary at the expense of the bulk.

\subsubsection*{The Solutions - Boundary Properties}

\begin{figure}[t!] 
  \centering
    \includegraphics[scale=0.6]{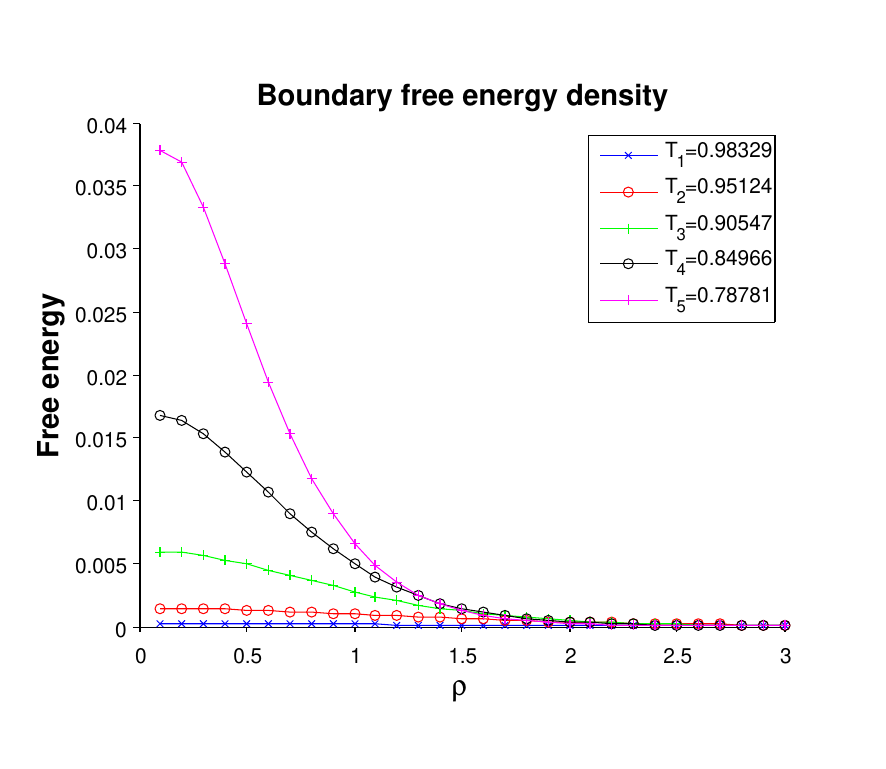}
    \caption{Bounary free energy density of the soliton for several values of the temperature relative to the critical temperature $T_{c}$. Notice the changes in the vortex profile as a function of temperature --- at low temperatures it is peaked near the vortex core while as the temperature increases it tends to become wider and more diffuse, tending to the homogeneous background at $T_{c}$.
}
\label{Energy_bound}
\end{figure}

\begin{figure}[t!] 
  \centering
    \includegraphics[scale=0.6]{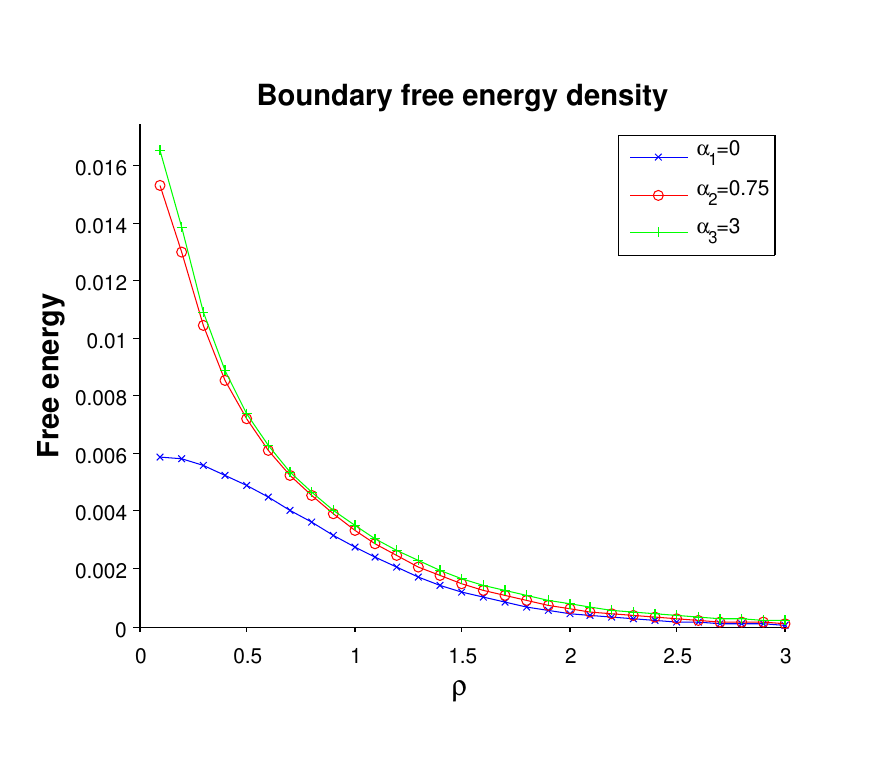}
    \caption{Profile of the boundary free energy density of the vortex, for various values of the boundary gauge coupling $\alpha$, at a temperature of $\simeq 0.91 T_c$. We note that the energy density as a function of $\alpha$ quickly begins to saturate.}
\label{Energy_alpha}
\end{figure}

The boundary free energy density can be found by the standard procedure of integrating radially the Euclidean on-shell action, and including both the counterterm action and the boundary Maxwell term (whose coefficient is $\alpha$). We now discuss  the boundary free energy  and its dependence on various parameters.

In figure \ref{Energy_bound} we display the boundary free energy density for several values of the temperature (at $\alpha=0.001$). At low temperatures (relative to the critical temperature) one sees that the vortex energy profiles are sharply peaked near $\rho=0$ and that, as one approaches the critical temperature, they flatten and broaden as the vortices begin to disperse. The vortex solutions merge with the homogeneous background at the critical temperature $T_{c}$.

Once we make the boundary gauge fields dynamical (i.e. turn on $\alpha$), the solutions significantly change and the free energy receives additional contributions from the boundary Maxwell action. In figure  \ref{Energy_alpha} we plot the total boundary free energy for various values of the coupling $\alpha$. We see that the energy density is that of a finite size lump, as expected, and that turning on $\alpha$ can be quite significant at the core of the vortex,  for the range of couplings displayed.

\subsubsection*{The Solutions - Dependence on Parameters}

\begin{figure}[h!]
  \centering
    \includegraphics[scale=0.5]{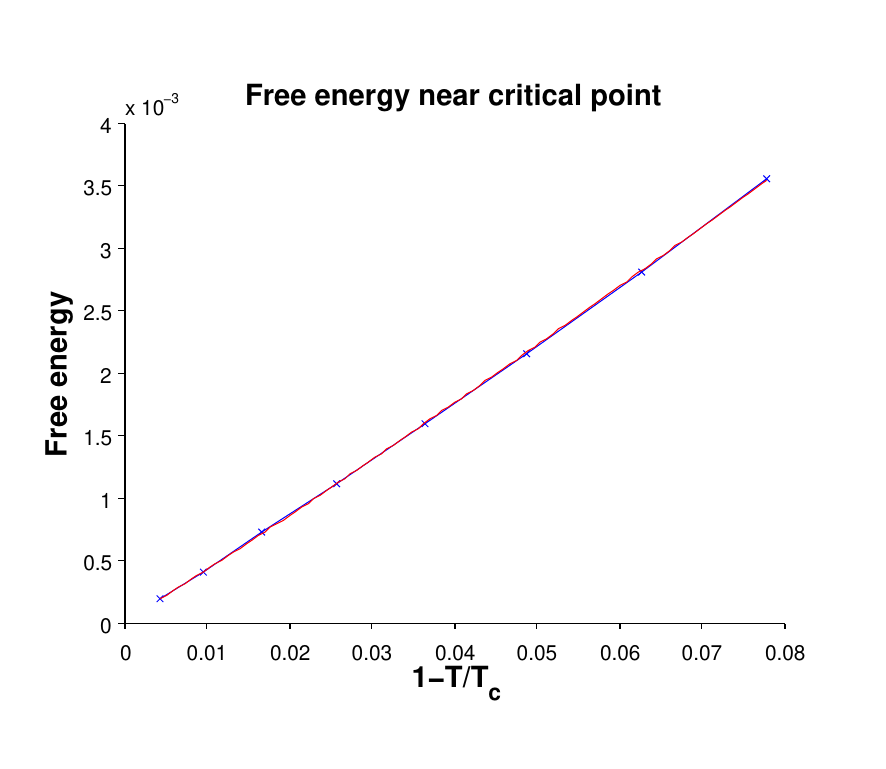}
     \includegraphics[scale=0.5]{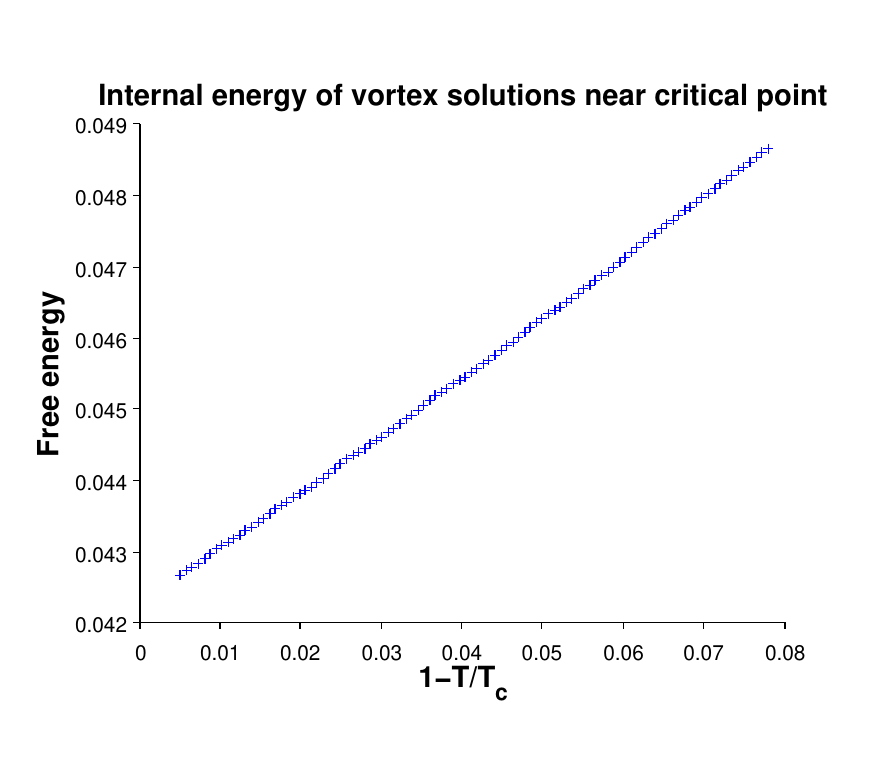}
    \caption{The total boundary free energy  and the string tension (internal energy) as a functions of temperature below the critical temperature. The nearly linear behavior is in agreement with the Landau Ginzsberg model of superconductivity near the critical point. However as these solutions are normalized with respect to the translationally invariant condensate the fact that the linear behavior continues to exist for the vortex solutions is noteworthy.}
 \label{FvsT}
\end{figure}

\begin{figure}[h!]
  \centering
     \includegraphics[scale=0.5]{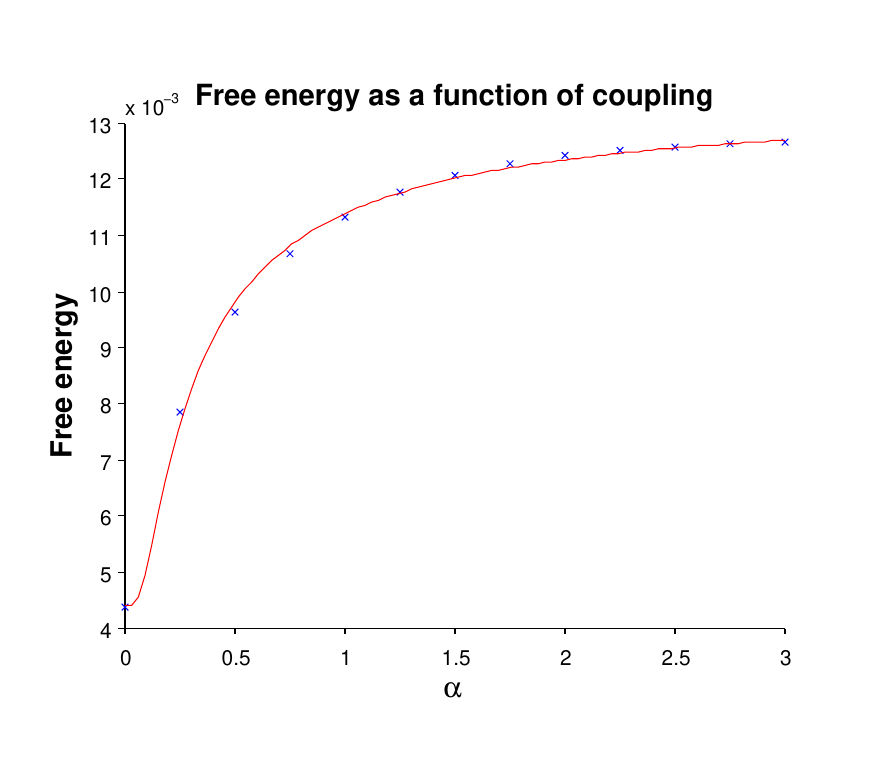}
    \caption{Boundary free energy density as function of the boundary gauge coupling $\alpha$. As expected the free energy remains bounded in the limit that $\alpha $ is taken to be small or large.}
 \label{freealpha}
\end{figure}

In order to display the dependence of the total boundary free energy on temperature, in figure  \ref{FvsT} we show the decrease in the free energy as we approached the critical temperature $T_{c}$ (found from examining the onset of the translationally invariant condensate). Fitting the curve to a function of the form \beq F = \alpha (1 -T/Tc)^\beta\eeq yields approximately $\alpha=0.0529,
\beta=1.0637$. In other words, up to numerical inaccuracies, the free energy of the soliton coincides with that of the translationally invariant (uncondensed) background at the critical point, and depends on temperature approximately linearly in the low temperature phase.

It is also interesting to examine the string tension (which corresponds to the internal energy), which quantifies the strength of magnetic confinement, as function of temperature. We exhibit that dependence in figure \ref{FvsT} as well, we see that the qualitative behavior is similar to that of the free energy. We note that the fact that the free energy goes to zero in an (approximately) linear fashion as one approaches the critical temperature ensures that the vortex solutions appear initially with some finite internal energy.

 In figure \ref{freealpha} we display the dependence of the total boundary free energy on the parameter $\alpha$. The dependence we find is intriguing: as we increase $\alpha$ (corresponding to decreasing the boundary gauge coupling) the free energy rises rapidly and eventually saturates, resulting in finite free energy difference between $\alpha=0$ and $\alpha \rightarrow \infty$. Fitting to a function of the form\footnote{This form is consistent with the existence of a perturbative expansion in the boundary gauge coupling.} \beq F= A \exp(-B/\alpha) + C \eeq yields approximately $A=0.009,
B=0.257, C= 0.004$. The exact interpretation of this result is unclear. We note that the large $\alpha$ limit corresponds to taking $e^{2}$ to zero (as the bulk coupling must be kept small in order for classical gravity to be valid.) Naively this would lead one to believe that the boundary term in the bulk gravity action, and the corresponding term in the field theory partition function, become free Maxwell theories. However as the boundary action serves to implement the boundary conditions for the bulk equations of motion and the gauge field in the field theory is an emergent component of a strongly coupled system this interpretation is probably incorrect. It would be interesting to investigate this issue further.

\subsubsection*{Electric Screening}

\begin{figure}[h!]
  \centering
    \includegraphics[scale=1]{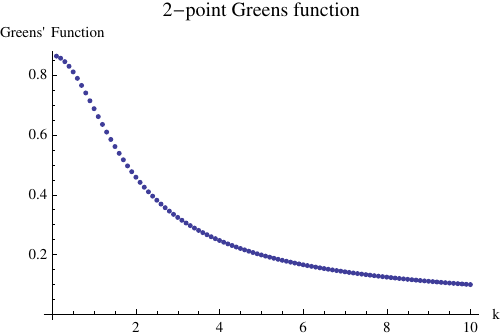}
    \caption{Zero frequency two point function of the boundary gauge field, in the limit of small momentum.}
 \label{screening}
\end{figure}

Finally, for the sake of completeness we comment on the behavior of the vacuum in the presence of electric sources. Instead of probing the response to those sources by calculating the Wilson line, it is simpler in our case to concentrate on the Green's function of  the boundary gauge field. While in the case of Dirichlet boundary condition the  Green's function encodes the optical conductivity, in the case of dynamical boundary conditions this encodes the electric response  of the system. In order to demonstrate the expected behavior of electric screening, we have to show that the static (zero frequency) long distance limit of the Green's function is gapped. We demonstrate the gap in figure \ref {screening} by displaying the low momentum limit of the zero frequency Green's function. This clearly stays bounded as we take the zero momentum (long distance) limit.

\subsubsection*{Conclusion}

In conclusion, in this section we constructed vortex solutions in the context of the holographic models of \cite{Hartnoll:2008vx}, for various values of the bulk and boundary parameters. These vortices signify the onset of local symmetry breaking. The imposition of the dynamical boundary conditions corresponds, via the prescription of \cite{Marolf:2006nd}, to a dual field theory with dynamical gauge field, with varying values of the boundary gauge coupling. This is evidenced, for example, by the fact that any boundary gauge transformation which is only a function of the boundary coordinates respects the dynamical boundary conditions on the bulk gauge field. We find that  in the spontaneously broken phase,  the symmetry breaking is manifested by  the existence of bulk vortex solutions with the expected properties  of superconducting vortices: there is no operator corresponding to a superfluid current on the boundary, and the vortex boundary energy is finite. In contrast, as found at \cite{Keranen:2009re}, the imposition of  Dirichlet boundary conditions leads to a theory which exhibits a global symmetry breaking and vortices with diverging energy, as expected in a superfluid. 

We expect that the criteria developed here for characterizing local and global symmetry breaking will have applications in other bottom-up holographic models. In particular it would be interesting to explore the applicability of these techniques to models of finite density QCD and color superconductivity.

\bigskip
\bigskip
\bigskip

\section*{Acknowledgements}

We thank Phillip Argyress, Carlos Hoyos, Andreas Karch, Rob Leigh, Gordon Semenoff  and Mark van Raamsdonk for useful conversations. We are especially grateful to Finn Larsen for collaboration in the initial stages of this project. We thank Compute Canada for use of the Westgrid facilities. The work is supported by discovery grant from NSERC, E.S. is partially supported by a National CITA Fellowship.
\bigskip\bigskip\bigskip

\section*{Appendix A: Vortex Solutions in Flat Spacetime}
The vortex solutions for the flat space Abelian Higgs model are well known. Here we provide a brief discussion following \cite{Tong:2005un}, for further information see also \cite{Rajaraman:1982aa}.

The action of the Abelian Higgs model is:
\begin{align}
L= -\frac{1}{4} F^{\mu \nu} F_{\mu \nu} + \frac{1}{2} (\nabla_{\mu} \Psi)^{*} (\nabla^{\mu} \Psi) -\frac{1}{4} \lambda(|\psi |^2 - F^2)^2
\nonumber
\end{align}
The constant $F$ is proportional to the VEV of the charged scalar field $\Psi$ breaking the $U(1)$ gauge symmetry. Finite energy configurations of the fields require that they obey the asymptotic conditions:
\begin{align}
|\Psi| \rightarrow \psi_0, \quad \nabla_{\mu} \Psi= (\partial_{\mu} \Psi - i q A_{\mu} \Psi) \rightarrow 0, \quad \text{as} \quad \mathbf{x} \rightarrow \infty
\nonumber
\end{align}
Requiring cylindrical symmetry, the form of the scalar and gauge fields at infinity are constrained to be:
\begin{align} 
\label{falloff}
& \Psi(r,\theta) \rightarrow \psi_0 \exp(i \alpha (\theta)) \nonumber \\
& A_{\mu} \rightarrow -\frac{i}{q} \frac{\delta_{\mu} \psi}{\psi}= \frac{1}{q r} \frac{d \alpha}{d \theta}, \quad \text{as} \quad r \rightarrow \infty
\nonumber
\end{align}
The winding of the phase, $\alpha$, at infinity is an integer, $s$, which is related to the quantised magnetic flux through the plane orthogonal to the magnetic vortices:
\begin{align}
s=\frac{q}{2 \pi} \int_0^{2 \pi} A_\theta r d \theta = \frac{q}{2 \pi} \oint \mathbf{A.} d \mathbf{l}= \frac{q}{2 \pi} \times \text{magnetic flux}
\nonumber
\end{align}
In order to see explicitly the localized nature of these vortices one is required to examine the equations of motion. Using the ansatz $A_r = A_0=0$, $A_{\theta}=A(r)$ and $\Psi(r, \theta)= \psi(r) \exp(i s \theta)$ we obtain the following equations:
\begin{align} 
&-\psi (r) \left[\left(\frac{s}{r}-q A(r)\right)^2 + \lambda (\psi(r)^2-F^2)\right]+\frac{\psi '(r)}{r}+\psi ''(r)=0 \nonumber\\
& \psi (r)^2 \left(\frac{s q}{r}-q^2 A(r)\right)-\frac{A(r)}{r^2}+\frac{A'(r)}{r}+A''(r)=0
\nonumber
\end{align}
The falloff conditions imply that $\psi(r) \rightarrow \psi_0 $ and $A(r) \rightarrow \dfrac{s}{q r}$ as $r \rightarrow \infty$. We may use this information to linearize the Maxwell equation for large $r$ by setting $\psi(r)=\psi_0$. Solving the resulting equation yields:
\begin{align}
A(r)\xrightarrow[r \rightarrow \infty ]{} \frac{s}{q r} + \frac{C_1}{\sqrt{r}} \exp{(-q F r)} 
\nonumber
\end{align} 
As the scalar and gauge field approach their asymptotic configurations at large $r$, we expect the action to be dominated by the potential term. Therefore in order to find the asymptotic behavior of the scalar field we examine perturbations of the potential. Requiring that the first derivative of the potential to vanish fixes the minimum at $\psi_0=\frac{F}{2}$. The fluctuations around this are of the form $F^2 \lambda \rho(r)$ where $\rho(r)$ is the deviation of the scalar field from $\psi_0$. Using this approximation for the potential in the scalar equation and setting $A(r)=\dfrac{s}{q r}$ we obtain:
\begin{align}
\psi(r)\xrightarrow[r \rightarrow \infty ]{} \psi_0 + C_2 \exp{(-\sqrt{\lambda} F r)} 
\nonumber
\end{align}
The localized nature of the vortex is evident from the exponential decay of the fields to their asymptotic values for large $r$. The full solution can be obtained by solving the equations numerically.

\bigskip\bigskip\bigskip

\section*{Appendix B: Details of the Numerics}

We solve the equations with the boundary conditions listed in section \ref{bound_cond} numerically using successive overralaxation (SOR) algorithm. To this end we discretize the equation on a lattice of finite mesh-size $h$ covering the domain of integration, such that  continuous spatial coordinates $(\rho,w)$ are represented by discrete pairs $(w_i,\rho_j)$, where $1\leq i \leq N_w, 1 \leq j \leq N_\rho$ are integers. We use a second order finite differencing approximation (FDA), where the derivatives are replaced with their finite differencing counterparts, e.g. $\partial_w R \rightarrow (R_{i+1,j}-R_{i-1,j})/2h, \partial_\rho R  \rightarrow (R_{i,j+1}-R_{i,j+1})/2h$ etc.  Following discretization, we thus obtain finite difference equations, at every mesh point, for each field.  We iteratively solve the entire system of algebraic equations using pointwise SOR starting with an initial guess for the fields, until a desired precision is achieved. Typically we initialize our scalar and $A0$ gauge field with the values of the homogeneous solution (found by solving the ODEs using shooting) and set the $A_\theta$ field to its expected asymptotic value of $s/q$. Along the horizon an initial guess is made for the scalar and $A_\theta$ fields which interpolates exponentially between the zero boundary condition at $\rho=0$ and the expected asymptotic values at the $\rho \rightarrow \infty$ boundary. We use similar SOR parameters for all fields. These are calculated at each step via Chebyshev iteration. In this iteration the spectral radius of the Jacobi iteration is chosen, for simplicity, to be that of the Laplace equation with Dirichlet boundary conditions, see  Section 19.5 of \cite{press1990numerical}  for further details and the algorithm.

While Dirichlet boundary conditions are implemented by assigning the fields their initial values throughout the relaxation procedure,  Neumann or Robin boundary conditions are updated after each iteration. We do this by using the backwards FDA derivative operators to update the boundary grid points based on the values calculated at the interior points. 
It turns out that at the horizon such a straightforward implementation of the regularity conditions (\ref{hor_con})  is numerically unstable. As these conditions relate the radial and tangential derivatives of the fields along the horizon they yield, upon discretization, a pair of coupled polynomial equations which relate the values of the fields at grid points in the near horizon region. Attempting to solve these polynomial equations to update the values of the boundary grid points after each iteration resulted in instabilities, which we attribute to the fact that the linearized scalar equation near the horizon is ill-posed (the effective mass terms and the elliptic operator have the same sign). The physical reason of this instability can be traced to the fact that the effective scalar mass in the near horizon region violates the Breitenlohner-Freedman bound, so that it triggers an instability and formation of a  condensate.

We found that a stable implementation of the constraint equations (\ref{hor_con}) is to evaluate all terms in the equation, except for the radial derivative, on the line of grid points just before the horizon, and to then use the FDA form of the radial derivative to extrapolate to the values of the fields on the horizon. This approach is consistent with the bulk equations of motion and identical to implementing the desired constraint equations when the continuum limit is taken (i.e. the limit in which the step size is taken to zero).

While our numerical lattice extends all the way from the horizon $w=1$ to the conformal boundary $w=0$, it covers only finite domain in the transverse direction $0<\rho<\rho_{cut}$. The truncation radius $\rho_{cut}$ is chosen such that our numerical solutions are altered by less than $0.01 \%$ when $\rho_{cut}$ is increased. Typically we use $\rho_{cut} \sim 100-120$. In addition, we checked that asymptotically our PDE solutions of the vortex configuration converged to the ODE solutions of the translationally invariant configuration at the transverse boundary to accuracies of $0.01 \%$ or higher.

Finally we discuss convergence of our finite-differencing numerical solutions. The rate of convergence is assessed based on the assumption that in the continuum limit, when the grid-size tends to zero, the discrete solution on the mesh $h$, designated $u_h$,  approaches the continuum solution, $u_*$, namely  $u_h = u_* + O(h^n)$. The power $n$ measures the rate of convergence. It can be calculated by running simulations with similar parameter settings on a sequence of meshes with decreasing mesh-spacings $h, h/2$ and $ h/4$, and computing $n=log_2 (u_h-u_{h/2})/(u_{h/2}-u_{h/4})$.  We found that the convergence rate in our case is very close to $n=2$ for the scalar and $A_\theta$ fields as expected for second order FDA, provided the numerical lattice is sufficiently dense to ensure we are in convergent regime. The convergence rate for for the $A_0$ field was seen to be somewhat lower at, $n \simeq 0.7$. Typical meshes that use to obtain results seen in \ref{Energy_bound} are of size $N_w \times N_\rho = 400 \times 1000$, which yields grid spacings of order $h_w \times h_\rho \simeq 0.0025 \times 0.1$.

\bigskip\bigskip\bigskip

\bibliographystyle{acm}
\bibliography{/Users/darrensmyth/Google_Drive_shared/Google_Drive/Thesis_writing/Thesis_attempts/myrefs_library}

\end{document}